\documentclass[prc,aps,nofootinbib,twocolumn,superscriptaddress]{revtex4}   
\usepackage{epsfig}  
\usepackage{graphicx}  
\usepackage{amssymb}
\usepackage{amsmath}
\usepackage{color}  

\usepackage{dsfont}

\begin{document}

\title{Transfer probabilities for the reactions $^{14,20}$O+$^{20}$O 
in terms of multiple time-dependent Hartree-Fock-Bogoliubov 
trajectories  }

\author{Guillaume Scamps}
\email{scamps@nucl.ph.tsukuba.ac.jp}
\affiliation{Center for Computational Sciences, 
University of Tsukuba, Tsukuba 305-8571, Japan}

\author{Yukio Hashimoto}                     
\email{hashimoto.yukio.gb@u.tsukuba.ac.jp}
\affiliation{Center for Computational Sciences, 
University of Tsukuba, Tsukuba 305-8571, Japan}

%


\begin{abstract}
 The transfer reaction between two nuclei in the superfluid phase is studied with the Time-dependent Hartree-Fock-Bogoliubov (TDHFB) theory. In order to restore the symmetry of the relative gauge angle a set of independent TDHFB trajectories is taken into account. Then, the transfer probability is computed using a triple projection method.
 This method is first tested to determine the transfer probabilities on a toy model and compared to the exact solution. 
 It is then applied to the reactions $^{20}$O+$^{20}$O and $^{14}$O+$^{20}$O in a realistic framework with a Gogny interaction.
\end{abstract}

%Use showkeys class option if keyword
\maketitle

%\section{Introduction}
%\label{intro} 

\textit{Introduction.} During the last decade, realistic calculations of time-dependent mean-field simulations have been developed including pairing correlations. 
The first time-dependent Hartree-Fock-Bogoliubov (TDHFB)  calculation was carried out by using a spatial lattice assuming spherical symmetry \cite{Ave08}.  
Then, three-dimensional (3D) calculations with  the BCS approximation for the pairing have been performed  \cite{Eba10,Eba14,Sca13}. 
The development of full 3D TDHFB calculations with no spatial symmetries have also been achieved \cite{Ste11, Bul16}. 
The hybrid basis method has permitted the conception of  TDHFB calculation with Gogny interactions \cite{Has12, Has16}.  
%Time-dependent density-matrix (TDDM) calculations that take into account the pairing and the collision term are also developed \cite{Toh02,Ass09,Toh16}.

Despite these progress, dynamical mean-field calculations suffer from the problem of the classical trajectory of collective variables. 
It is for example impossible to describe the tunneling in fusion or fission reactions.
For fission, the Time-dependent Generator Coordinate Method (TDGCM) \cite{Gou05,Reg16,Zde17} solves partially this problem but assumes a collective motion in a space of adiabatic basis. % an adiabatic path for fission. 
On the other hand, for fusion tunneling, the Density-constraint Hartree-Fock \cite{Uma06,Uma15} is restrained to a single energy-dependent fusion path.

Another example of limitation of time-dependent mean-field calculation is the lack of fluctuation, due to the consideration of only one quasi-particle vacuum state to describe a reaction. 
With the aim of curing the problem and getting over the limitations, the time-dependent random-phase approximation (TDRPA) \cite{bal84,bro08,Sim11BV,sim12b} or the stochastic mean-field (SMF) \cite{Ayi08,Ayi09,Was09,Yil11} are interesting directions. 
We can note in particular, the recent article \cite{Tan17}, which describes the fission with an ensemble of TDHF+BCS trajectories with different initial densities determined randomly.
 The observables after the scission are computed as the average value over the statistical ensemble of trajectories.
  In that case, the quantal fluctuations are supposed to behave like statistical fluctuations.

In the present approach, we would like to consider the evolution of a state that is a mixture of several TDHFB states (trajectories). 
In a precedent contribution, we showed that the fusion dynamics \cite{Has16} changes with the initial relative gauge angle. 
In addition, being analogous with the well known Josephson effect, the nucleus-nucleus potential contains a non-negligible pairing contribution that varies with the gauge angle of the two fragments.
 This has been confirmed for heavier nuclei \cite{Mag16,Sek17}, where it is shown that the solitonic excitation due to the gauge angle difference can change the barrier height by several MeV. 

As is discussed in Ref.  \cite{Bul17}, when we treat a case in which two superfluid objects are close to each other and exchange 
pairs of particles between them, it is necessary to project on the initial particle difference number. 
This gives us an ideal test case to investigate the evolution of a projected states in which various gauge angles are mixed at the initial time.
%mixing initially different gauge angles. 

The projection method to compute the transfer probability in time-dependent mean-field calculation has been proposed in Ref. \cite{Sim10}.
It has been used in order to make realistic prediction of cross-section in multi-nucleon transfer reactions \cite{Sek13}.
The method has been generalized for calculation where one of the fragment is in the superfluid phase \cite{Sca13}.
This method can be used  to understand the excitation energy in transfer channel \cite{Son15,Sca17}.
In this contribution, we use the mixing of TDHFB trajectories to compute the transfer probabilities in reactions where both fragments are initially superfluid.

%In this article, the projection method is first described and tested with a simple toy model in the Sec. \ref{method}. 
%And then it is applied to a realistic model in Sec \ref{realistic}. We summarize the paper in Sec.  \ref{Summary}
In this article, the projection method is first described and tested with a simple toy model
and then it is applied to a realistic model.  

%\section{ Method }
%\label{method}

\textit{Method.} Here, we test the projection method by using a simple toy model. 
We have two systems, left (L) and right (R), which are composed of $\Omega$ levels, each of which is doubly degenerate. 
The two systems are separately initialized  with the following Hamiltonian,
\begin{align}
\hat H_{0} =  \sum_{i>0}  h_i (\hat a^{\dagger}_{i} \hat a_{i} + \hat a^{\dagger}_{\bar i} \hat a_{\bar i}   )  +  G \sum_{i \ne j } \hat a^{\dagger}_{ i} \hat a^{\dagger}_{\bar i}  \hat a_{ \bar j} \hat a_{ j}, 
\end{align}
with mean-field energy $h_i$ = $i$ Mev with $i=1$ to $\Omega$. 
It is assumed in this model that all the particles are paired.
The HFB method is used to determine the ground states of the two systems with a constraint to have $N^0_{L}$ ($N^0_{R}$) particles in the left (right) sytem, respectively. 
Then, to simulate a collision, an HFB state $|\phi(t=0) \rangle$ of a set of the two systems is constructed from the two quasi-particle vacuum states.

From this HFB state that breaks the exact conservation  of the particle number, a projected state is initially created,
\begin{align}
| \Psi(t=0) \rangle =  \hat{P}_{L}(N_L^0) | \phi(t=0) \rangle , \label{eq:proj_left}
\end{align}
where $\hat{P}_{L}(N_L^0) $ is a projector to a state with $N_L^{0}$ particles in the left sytem, 
\begin{align}
	\hat P_{L}(N_L^0) = \frac{1}{2 \pi} \int_0^{2 \pi}  e^{i \varphi \left( \hat N_{L} - N_L^0 \right) } d\varphi.
\end{align}
This projection is made in order to restore the symmetry with respect to the relative gauge angle. 
We chose arbitrarily to project initially on the left, but a projection on the right or on the initial difference of number of particles would lead to the same results.
The integral is discretized with the Fomenko method \cite{Fom70}, with $M$ points,
\begin{align}
	|  \Psi(t=0) \rangle =  \frac{1}{ M }   \sum_{n=1}^{M}  e^{- i \frac{2 n \pi}{M} N_L^0 } | \phi_n (t=0) \rangle , \label{eq:proj_form}
\end{align}
and 
\begin{align}
	|  \phi_n(t=0) \rangle =   e^{i \frac{2 n \pi}{M} \hat N_L } | \phi (t=0) \rangle.
\end{align}
The initial state is then a mixing of $M$ HFB states with different  initial relative gauge angles.
In the toy model, the transfer of particles is induced by a time dependent coupling between the two systems that is assumed to be gaussian as a function of time.
Then, the total Hamiltonian of the system is written as the sum of the initial left  $\hat H_{0}^{(L)}$ and right $\hat H_{0}^{(R)}$ hamiltonians and a coupling term,
\begin{align}
\hat H(t) = & \hat H_{0}^{(L)} +  \hat H_{0}^{(R)}  \nonumber \\&+ V(t) \sum_{i \in L ; j \in R }  \left(   \hat a^{\dagger}_{ i} \hat a^{\dagger}_{\bar i}  \hat a_{ \bar  j} \hat a_{ j}  +   \hat a^{\dagger}_{ j} \hat a^{\dagger}_{\bar j}  \hat a_{ \bar i} \hat a_{ i}  \right),
 \label{eq:hamiltonian_toy}
\end{align} 
with 
\begin{align}
V(t) = V_0 e^{-at^2}.
\end{align}

The natural way of performing this calculation would be  to derive the equation of motion from the minimization of the action of the evolution of the projected system. Nevertheless, this method would require to compute the matrix elements of the Hamiltonian matrix, which would lead to spurious results with the use of the density-dependent effective interaction \cite{Sim14}.

Then, in this calculation,  we assume that the form of the mixing (Eq. \eqref{eq:proj_form}) does not change as a function of time,
\begin{align}
	|  \Psi(t) \rangle =  \frac{1}{ M }   \sum_{n=1}^{M}  e^{- i \frac{2 n \pi}{M} N_L^0 } | \phi_n (t) \rangle, 
\end{align}
 and  that every HFB state evolves independently with the following TDHFB equation of motion,
\begin{align}
  i \hbar \frac{\partial}{\partial t} 
    \left(
    \begin{array}{c}
                U_k(t) \cr
                V_k(t)
             \end{array}
             \right)
     = {\cal H}_k(t)
     \left( 
    \begin{array}{c}
                U_k(t) \cr
                V_k(t)
             \end{array}
             \right), 
\end{align}
with
\begin{align}
  {\cal H}_k(t)= \left( 
              \begin{array}{cc}
               h- \delta\lambda_{L,R} -\epsilon_k & \Delta \cr
               - \Delta^{*} & -h^* + \delta\lambda_{L,R}-\epsilon_k
             \end{array}             
             \right).
             \label{test}
\end{align}
with $\epsilon_k(t)$ the quasi-particle energy.

%The quasi-particle energy $\epsilon_k(t)$ is computed as, 
%
%\begin{align}
%\epsilon_k(t) = \left( \frac{}{} U^\dagger_k(t)  \quad V^\dagger_k(t) \right) {\cal H}(t) \left( 
%    \begin{array}{c}
%                U_k(t) \cr
%                V_k(t)
 %            \end{array}
%             \right).
%\end{align}
%

\begin{figure}[htb]
\begin{center}
\includegraphics[width= 0.95 \linewidth]{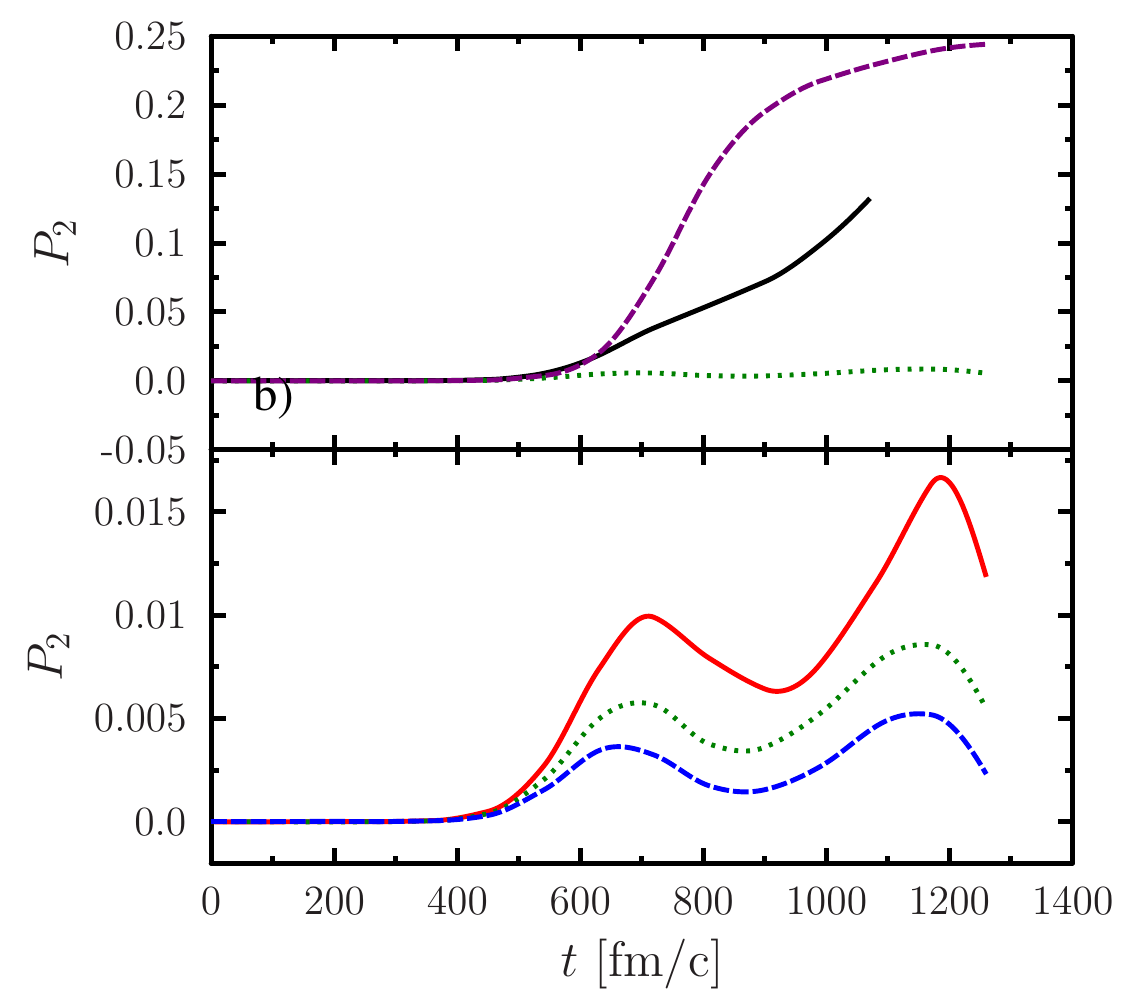}
\end{center}
\caption{ Pair transfer probability as a function of time  for the reaction $^{20}$O+$^{20}$O.
a) TDHFB calculation without center-of-mass shift (purple dashed line), without the chemical potential and quasi-particle energy correction (solid line), and the correct calculation at the center-of-mass energy $E_{cm}=9.21$ MeV (green dotted line).
b) Corrected calculation at the center-of-mass energy $E_{cm}=9.31$ MeV (red solid line), 
$E_{cm}=9.21$ MeV (green dotted line) and $E_{cm}=8.91$ MeV (blue dashed line).}
\label{fig:P2_fct_t}
\end{figure}

The $ \delta\lambda_{L,R}(t) $ variable is defined as the difference $\lambda_{L,R}(t)-\lambda_{L,R}(t=0)$.
The chemical potential $\lambda_{L}$ ($\lambda_{R}$) in the left  (right) subspace is introduced in order to avoid phase rotation  among TDHFB states  \cite{Sca17_proc},
\begin{align}
\lambda_{L,R} = \rm{ Real} \left( \frac{ \sum_{ i \in L,R}  \kappa_{i\bar{i}} M_{i\bar{i}} }{ \sum_{ i \in L,R} \kappa_{i\bar{i} } } \right),
\end{align}
with the $M_{i\bar{i}}$ matrix,
\begin{align}
 M_{i\bar{i}} = \frac1{\kappa_{i\bar{i}}} \left[ \frac12  h \kappa + \kappa h^* - \Delta \rho^*  + (1 - \rho) \Delta  \right]_{i\bar{i}} .\label{eq:mmatrix}
 % - i \frac{d \kappa}{d z}  \frac{v_{L,R}}{ 2 \kappa }, 
\end{align}
The $ \delta\lambda_{L,R}(t) $ is equal to  $ \delta\lambda_{L}(t) $ for $z<0$ and $ \delta\lambda_{R}(t) $ for $z>0$. 
The phase rotation in the process of the TDHFB evolution would induce a spurious time dependence of the results after the transfer as we can see in Fig. \ref{fig:P2_fct_t}.

The main observables of interest here are the transfer probabilities.
To calculate them, we use a projection method \cite{Sim10} to determine the probability $P_{L,R}(N,t)$ to obtain $N$ particles in the left or right part of the system at the time $t$,
\begin{align}
P_{L,R}(N,t) = \langle \Psi (t) |  \hat{P}_{L,R}(N) | \Psi (t) \rangle.
\end{align}
As shown in Ref. \cite{Sca13},  in a theoretical framework in which the many-body states mix different total number of particles, 
it is necessary to select the components that contain the  good total number of particles. 
%Here, we project initially on the good initial number on the left (see Eq. \eqref{eq:proj_left}) and at the end on the total number $N_{\rm tot.} = N_L + N_R$, 
Here, we have projected initially on  the good initial number on the left (see Eq. \eqref{eq:proj_left}). In addition, we needed to project on the total number of particles $N_{\rm tot.} = N_L + N_R$ before making a measurement on the system,
% the projection on a state with the total number $N_{\rm tot.} = N_L + N_R$ is followed 
%by the projection on a state with the good initial number in the left or right 
%system (see Eq. \eqref{eq:proj_left}),
%
\begin{align}
P_{L,R}(N,t) &= \frac{1}{\cal N} \langle \Psi(t) |  \hat{P}_{L,R}(N)  \hat{P}( N_{\rm tot.} ) | \Psi (t) \rangle, \\
&=  \frac{1}{M^2 \cal N} \sum_{n,m} e^{ i \frac{ 2 (n-m)  \pi }{M} N_L^0 } \nonumber \\ 
&\quad \quad \langle \phi_n (t)  |  \hat{P}_{L,R}(N)  \hat{P}( N_{\rm tot.})   | \phi_m(t) \rangle. \label{eq:proj_x3}
\end{align}
Here, $\cal N$ is the norm,
\begin{align}
{\cal N} =  \langle \Psi(t) |   \hat{P}( N_{\rm tot.} ) | \Psi (t) \rangle,
\end{align}
and $\hat{P}( N_{\rm tot.} )$ the particle number projector on the total number $N_{\rm tot.} = N_{L}^{0} + N_{R}^{0}$.

Note that the projection $\hat{P}( N_{\rm tot.} )$ can also be made at the initial time and so added in the eq. \eqref{eq:proj_left}. But these would not change the evolution since the TDHFB equation are independent of a  rotation of the gauge angle associated to the total number of particles. 

\begin{widetext}

To compute the overlaps in Eq. \eqref{eq:proj_x3}, we develop, respectively, the projector $\hat{P}_{L,R}(N)$ and $\hat{P}( N_{\rm tot.})$   as integral  with respect to the gauge angle $\varphi_1$ and $\varphi_2$.
Then, we use the Pfaffian method \cite{Ber12} to compute,
\begin{align}
	\langle \phi_n(t) |   e^{ i \varphi_2 \hat N_{L,R}  }  e^{ i \varphi_3 \hat N_{\rm tot}  }       |   \phi_m (t) \rangle& = 
		(-1)^{n_{\rm qp}} \frac{\det C^* \det C'}{\prod_\alpha^n v_\alpha v'_\alpha} \nonumber \\
		& \times {\rm pf} \left[     
		\begin{array} {cc}
			V^TU & V^T e^{i \varphi_3}(1+\Theta_{L,R}({\vec r}) (e^{i \varphi_2}-1) ) V'^* \\
			-V'^\dagger  e^{i \varphi_3}(1+\Theta_{L,R}({\vec r}) (e^{i \varphi_2}-1) ) V & U'^\dagger V'^*
		\end{array}
		    \right]. \label{eq:overlap}
\end{align}

\end{widetext}

%%%

Here, $C$, $v_\alpha$, $V$ and $U$  correspond  to the bra $ \langle \phi_n(t) | $ 
while the $C'$, $v'_\alpha$, $V'$ and $U'$  refer to the ket  $|   \phi_m (t) \rangle $. 
The matrices $C$ and $C'$ are obtained from the Bloch-Messiah decomposition \cite{RS}. 
 $v_\alpha$ and $v_{\alpha}'$ are the occupation numbers in the canonical basis and $n_{\rm qp}$ is the number of quasi-particle states. 
The Heaviside function $\Theta_{L}({\vec r})$ ($\Theta_{R}({\vec r})$) is equal to one in the left (right) part of the system, respectively. 

\begin{figure}[htb]
\begin{center}
\includegraphics[width= 0.9 \linewidth]{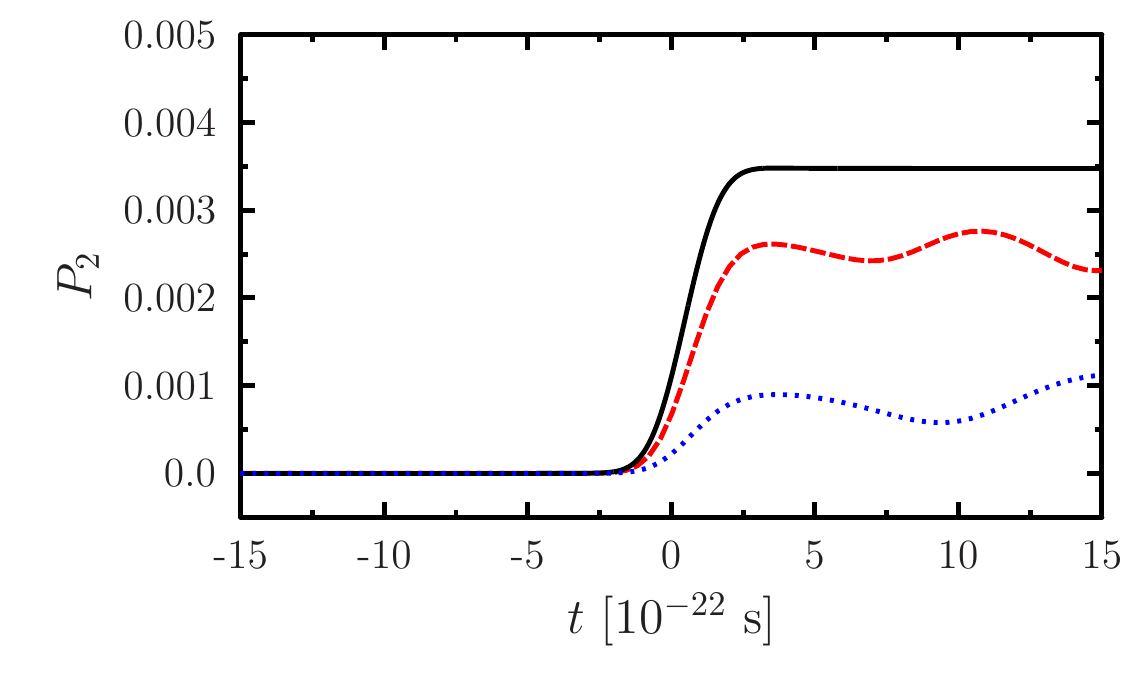}
\end{center}
\caption{ Pair transfer probability as a function of time in the toy model. 
Calculation by the present method with (red dashed curve) and without (dotted blue line) an effective interaction  are  compared to the exact solution (black solid curve).   }
\label{fig:res_comp_exact}
\end{figure}

To test our present method, we make a comparison with the exact solution of the many body problem. The exact solution is obtained by first, diagonalization of the Hamiltonian in the space of all the possible configurations and then solving the time-dependent Schr\"odinger equation. 
In the Hamiltonian Eq. \eqref{eq:hamiltonian_toy}, we use the following parameters, $\Omega=4$, $N_L^0=N_R^0=4$, $G=-1$ MeV, $V_0=-0.03$ MeV, and $a=0.3 \times 10^{44}s^{-2}$. 
The calculation is carried out between the time $t_{i} = -15\times 10^{-22}$ s until $t_{f} =15\times 10^{-22}$ s. 
Two calculations are carried out, one with the above parameters and one where
the interaction constants $G$ and $V_0$ are multiplied by 1.475 in the HFB and TDHFB calculation in order to obtain an effective interaction that reproduces the initial exact energy.

The use of a gaussian type coupling as a function of time allowed us to mimic the reactions. 
Initially at $t = t_{i}$, there is no coupling between the two systems. 
Then, around $t = 0$ the two systems interact via the pairing interaction and can exchange pairs. 
At the final time $t = t_{f}$, the systems are isolated form each other. 
The results as a function of time are shown in Fig. \ref{fig:res_comp_exact}. 
The transfer takes place in the vicinity of $t = 0$. After $t = 3 \times 10^{-22}$s, the transfer probability is constant in the exact case 
while small oscillations occur in the TDHFB result. 
% about the constant value in the exact case 
This phenomenon is due to the choice of the approximations made in the calculation which is based on an assumption that the TDHFB trajectories in the state $|\Psi(t)\rangle$ are independent of each other.
%the independence of the TDHFB trajectory, this assumption does not ensure the correct behavior (to be constant with respect to the time) 
%in the transfer probability after the interaction. 
Nevertheless, the averaged value of the transfer probability as a function of time is close to the exact solution with the use of the effective interaction.
As we can see in Fig. \ref{fig:res_comp_exact}, the use of the effective interaction increases the pair transfer by a factor 3.
The use of an effective interaction is justified here because the evolution of the system is made by a set of TDHFB calculation.
In each trajectory,  the transfer of pair by Josephson effect is governed by the pairing field $\Delta$. To obtain a correct result, the interaction strenght should be increased in order to compensate the lack of correlations due to the HFB approximation. This is simply realized by adjusting the interaction strength by using the same factor as is used to reproduce the ground state energy which also varies with $\Delta$.
%To adjust the interaction with the same factor as applied to reproduce the ground state energy seems to be a good approximation. 

\begin{figure}[htb]
\begin{center}
\includegraphics[width= 0.9 \linewidth]{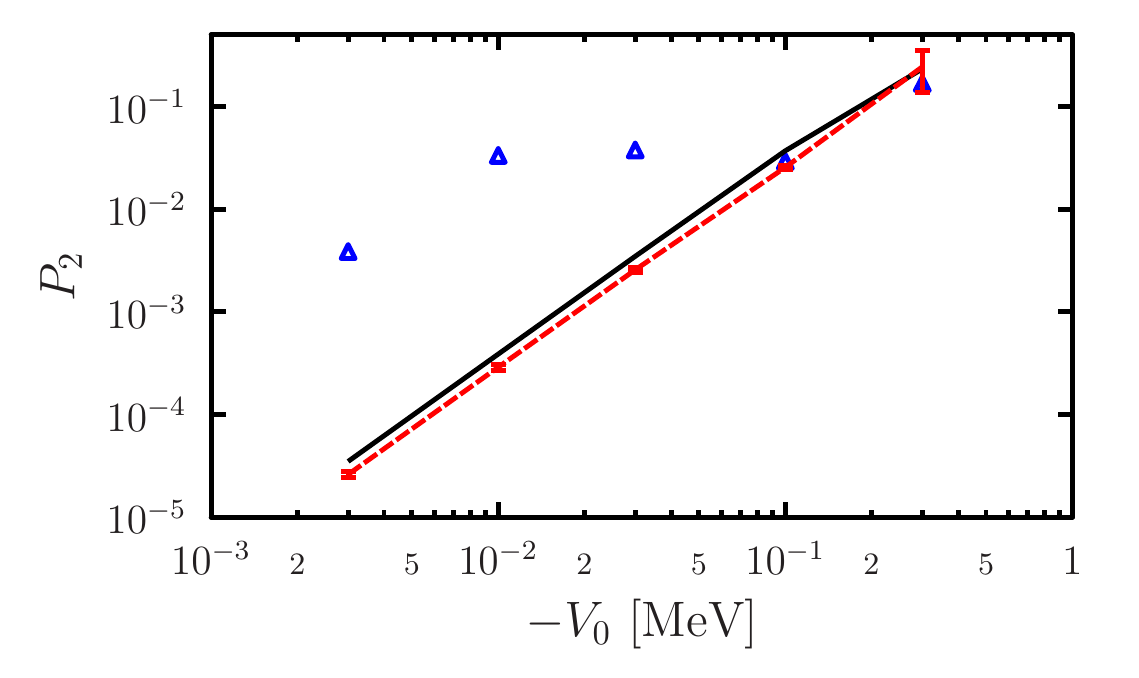}
\end{center}
%\caption{ Same as Fig. \ref{fig:res_comp_exact} as a function of the transfer coupling. }
\caption{Pair transfer probability as a function of the strength $V_0$ of the transfer coupling in the Hamiltonian \eqref{eq:hamiltonian_toy}. 
Calculation by the present method with the effective interaction (red dashed curve) is compared to the exact solution (black solid curve). The blue triangles represent the results obtained without the chemical potential and quasi-particle energy correction at an arbitrary time of 50$\times10^{-22}$ s.  The error bars on the red curve represent the  standard deviation of the probabilities as a function of time after the reaction.  } 
\label{fig:P2_fct_V_toy_model}
\end{figure}

With the purpose of illustrating the relation between the averaged value of the transfer probability in the TDHFB case and that in the exact case, 
we plot the pair transfer probability as a function of the strength of the transfer coupling in Fig. \ref{fig:P2_fct_V_toy_model} .
 The error bars represent the standard deviation of the probabilities as a function of time after the reaction (here for time t $>3 \times 10^{-22}$ s ).
 We  see that the TDHFB results reproduce well the exact solution when the interaction strength is varied in several orders of magnitude. To show  the importance of the prescription concerning the modification of the TDHFB equation (Eq. \eqref{test}) we show pair transfer probability at a large time  of 50 $\times10^{-22}$ s. We see that the transfer probabilities are over-estimated by two orders of magnitudes compared with the results of the calculations with the present prescription as well as the exact cases.
 
 In this way, the results of the calculation in the toy model illustrates that the present method works well 
 and we could expect that it has a predictive power in the realistic  TDHFB calculations.

%\section{realistic calculation}
%\label{realistic}

\textit{Realistic calculations.} To make the calculation with a realistic interaction, we use the Gogny TDHFB code that use an hybrid basis. 
The quasi-particle vacuum states  of the two Oxygen nuclei are separately initialized  and then put in the same lattice with a boost.
 The method  to form the initial HFB state $| \phi(t=0) \rangle$ is described in Ref. \cite{Has16}.

The Gogny D1S effective interaction was used in the calculations presented below.
The projector  in Eq. \eqref{eq:proj_form} is discretized with eight sampling points $M = 8$. 
Using the symmetry property, only the first four gauge angles are used. 
We calculated a test case with the sampling points $M = 16$ and found that there is practically no difference 
in the numerical results below between the cases with $M = 8$ and $M = 16$. 
%A test calculation with $M$=16 has been tested and show no differences with $M=8$.
The projectors at the final time are discretized with 16 points.

\begin{figure}[!ht]
\centering
\includegraphics[width= 0.9 \linewidth]{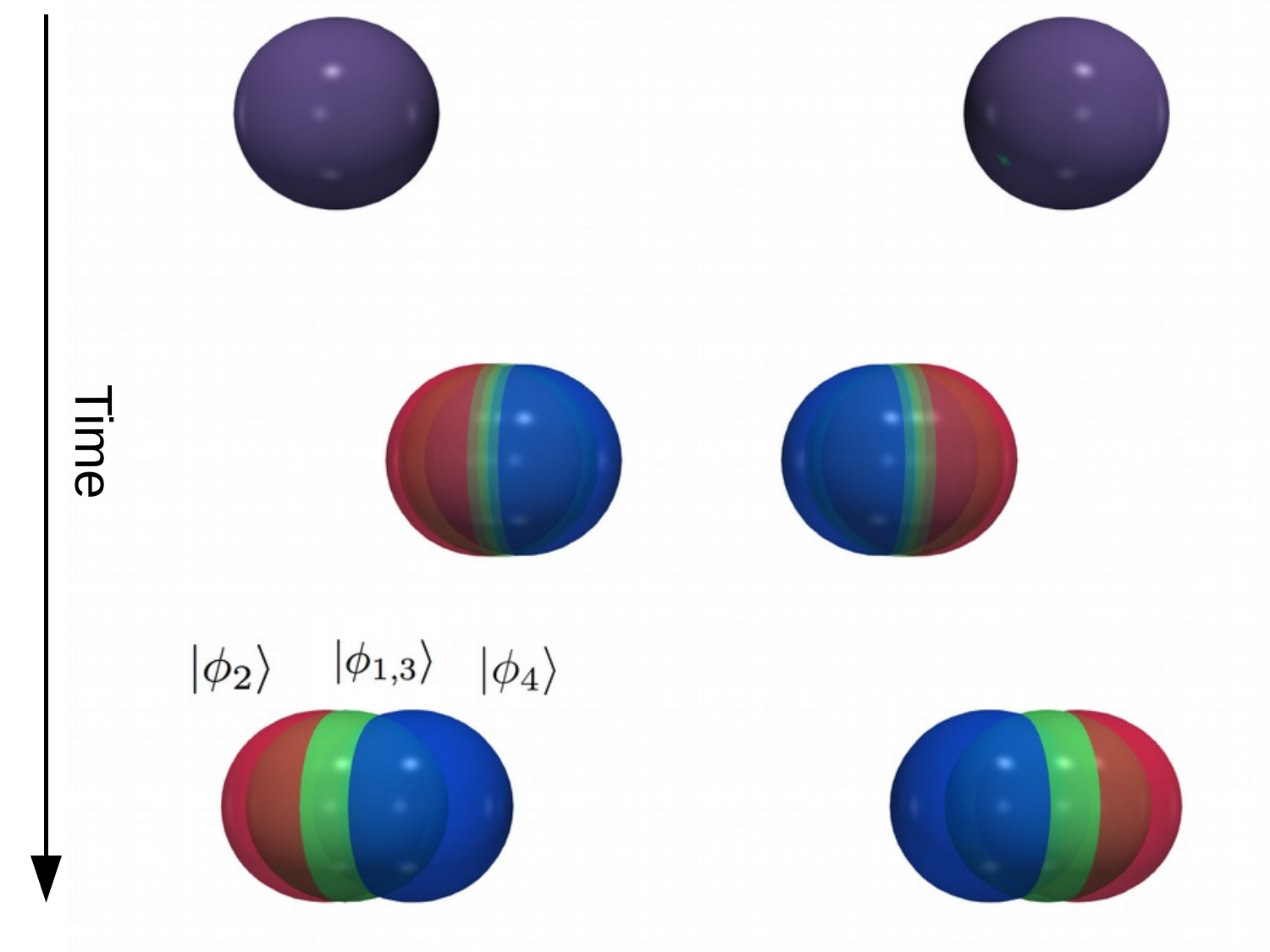}
\caption{ Superposition of the isodensity surfaces at $\rho$=0.8 fm$^{-3}$ as a function of time, 
for the reaction $^{20}$O+$^{20}$O with the center-of-mass energy $E_{\rm cm}$=9.31 MeV. 
The red and blue surfaces are respectively computed from the state $|\phi_2 \rangle $ and $|\phi_4 \rangle $  while the green surface is computed from the states  $|\phi_1 \rangle $ and $|\phi_3 \rangle $.}
\label{fig:film}      
\end{figure}

The evolution of TDHFB trajectories with four different gauge angles is shown on Fig. \ref{fig:film}. 
After the reaction, there is a splitting among the trajectories that is due to the dependence of the nucleus-nucleus potential on the initial gauge angle \cite{Has16}. 
The trajectory of $| \phi_2 \rangle$ corresponding to a relative gauge angle of $\varphi = 90^{\circ}$ is the one with the highest barrier 
and so the smallest contact time. 
The opposite case is $\varphi = 0^{\circ}$ ($|\phi_4\rangle$) with the largest contact time. 

\begin{figure}[htb]
\begin{center}
\includegraphics[width= 0.9 \linewidth]{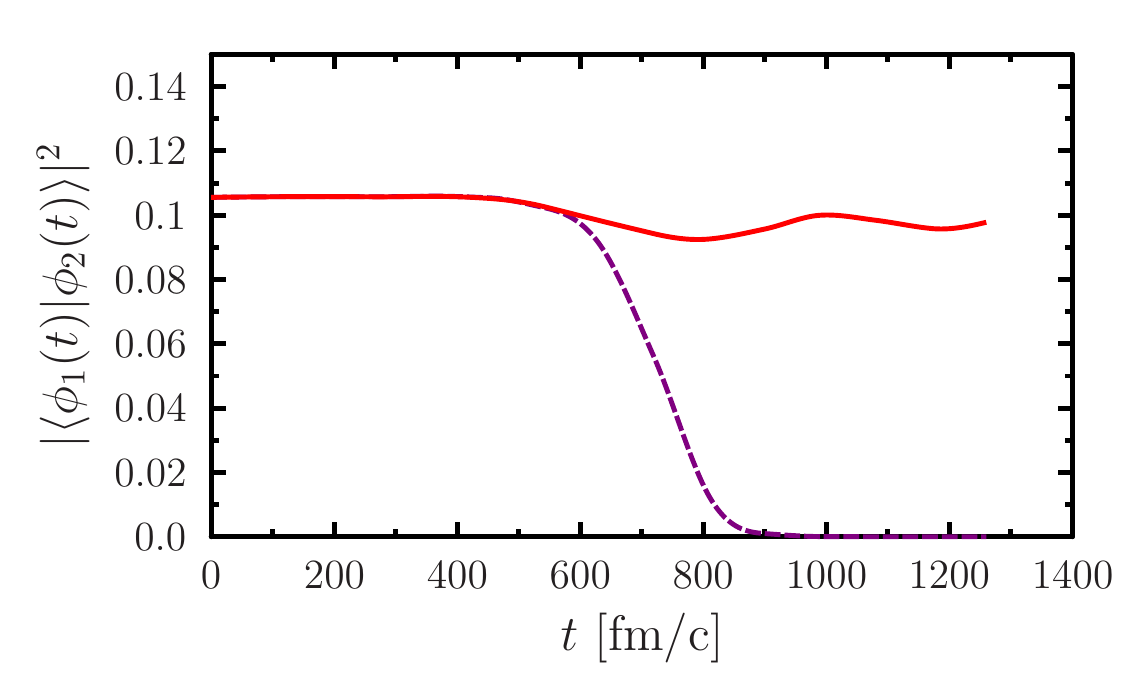}
\end{center}
\caption{ Norm of the overlap between the state 1 and 2 (respectively $\varphi=$0 and 45 degrees) as a function of time with the center-of-mass shift (red solid line) and without (purple dashed line). }
\label{fig:over_fct_t}
\end{figure}

This effect which gives rise to the splitting among the trajectories, changes the overlap as a function of time. As we can see on Fig. \ref{fig:over_fct_t}, the overlap $|\langle \phi_1(t)|\phi_2(t)\rangle|^2$ is constant until the contact, then the overlap drops to zero because the trajectories diverge. 
So the probabilities in a similar way as the phase rotation change as a function of time.
We can see in Fig. \ref{fig:P2_fct_t}, that the direct application of the method  (purple dashed line) leads to a very large pair transfer probability that is not stabilized after the reaction. 
To solve this problem, we introduce a prescription to  compensate the dispersion of the trajectories. We artificially shift the position of each fragments of the ket $ |  \phi_m (t) \rangle  $ to have the same center of mass  of the bra $ \langle \phi_n (t) | $,
%the center of mass of each fragments of the ket $ |  \phi_m (t) \rangle  $ to maximize the overlap with the bra $ \langle \phi_n (t) | $,
%
\begin{align}
U' ( \vec r ) \rightarrow U'' ( \vec r ) = U' ( \vec r  - \vec d_{L,R} ), \label{eq:transU}  \\
V' ( \vec r ) \rightarrow V''( \vec r ) = V' ( \vec r  - \vec d_{L,R} ), \label{eq:transV} 
\end{align}
with $\vec d_{L,R}$ the difference of position between  the bra and the ket in the left or right part of the lattice,
$\vec d_{L,R}  = {\vec R}_{L,R} -  {\vec R'}_{L,R} $. 
${\vec R}_{L,R} $ (${\vec R'}_{L,R} $) is the center-of-mass position in the left or the right of the system associated to the bra and ket, respectively. With this prescription, we ensure that there is no diminution of the overlap after the separation of the fragments.

It is also necessary to take into account this shift of the position in the calculation of the chemical potential. 
The matrix $M_{i\bar{i}}$ in Eq. \eqref{eq:mmatrix} is changed into, 
\begin{align}
 M_{i\bar{i}} = \frac{1}{\kappa_{i\bar{i}}} \left[ \frac12  h \kappa + \kappa h^* - \Delta \rho^*  + (1 - \rho) \Delta    - i \hbar \frac{d \kappa}{d z}  \frac{v_{L,R}}{ 2  } \right]_{i\bar{i}} , 
\end{align}
with $v_{L,R}$ the velocity of the left or right fragment in the $z$ direction (the reaction axis). We can see in Fig. \ref{fig:over_fct_t}, that with this correction the overlap remains almost constant after the reaction. 

The results of the corrected neutron pair transfer probability are shown as a function of time in Fig. \ref{fig:P2_fct_t}. 
Despite the efforts to  conserve the probability constants after the reaction, small oscillations arise after the reactions.  
Nevertheless, we expect that the average behavior will fairly reproduce the exact pair transfer probability as it happens in the toy model. 
Here, let us point out another numerical technique of introducing the chemical potential in Eq. (\ref{test}). 
The calculation without the chemical potential and quasi-particle energy is shown also on Fig. \ref{fig:P2_fct_t}. 
As is clearly seen, the transfer probability is not stable after the reaction, illustrating the improving effects 
of the chemical potential  and quasi-particle energy terms in Eq. (\ref{test}).

%The justification of this equation of motion can be found in \cite{Sca17_proc} as well as a test in a toy model. %Note that the second term is due to the translation that is done in the computation of the overlap (Eq. \eqref{eq:transU} and \eqref{eq:transV} ). 

\begin{figure}[htb]
\begin{center}
\includegraphics[width= 0.9 \linewidth]{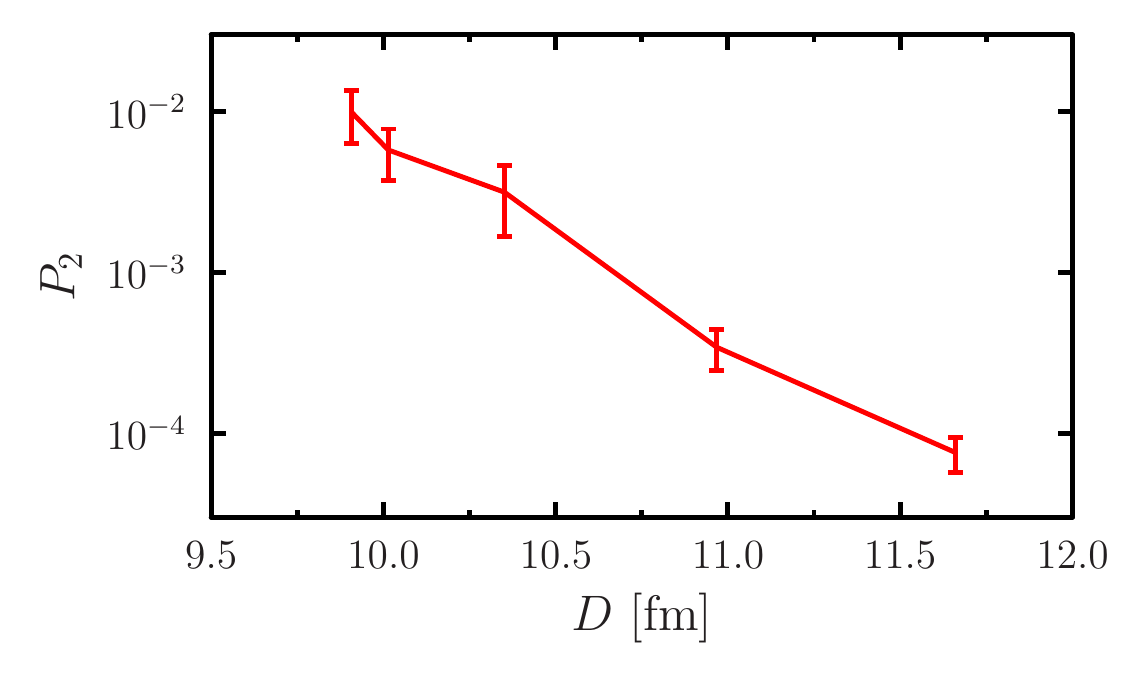}
\end{center}
\caption{ Neutron pair transfer probability as a function of the distance of closest approach for the reaction $^{20}$O+$^{20}$O. 
}
\label{fig:P2_fct_D_O20_O20}
\end{figure}

\begin{figure}[htb]
\begin{center}
\includegraphics[width= 0.9 \linewidth]{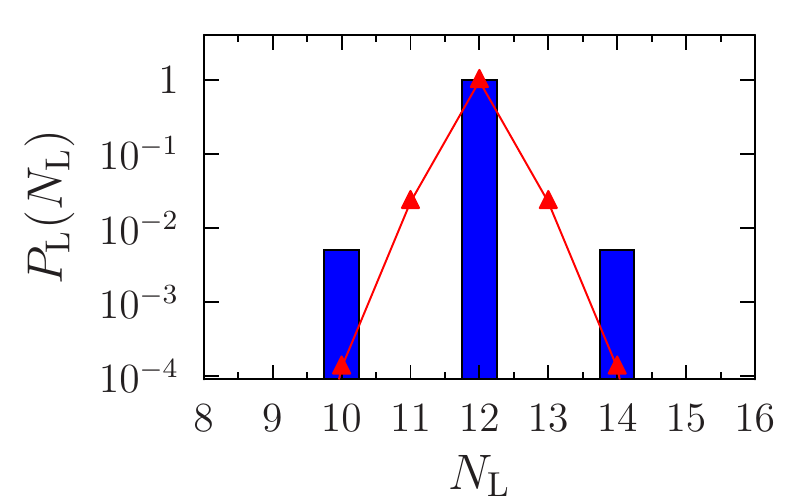}
\end{center}
\caption{ Probabilities to obtain $N_{\rm L}$ neutrons on the left part ($Z<0$) of the lattice after the collision computed with pairing (blue bars) and without pairing (red triangles) for the reaction $^{20}$O+$^{20}$O at  the center-of-mass energy $E_{cm}=9.31$ MeV.  } 
\label{fig:distr}
\end{figure}

We show the neutron pair transfer probability for the reaction $^{20}$O+$^{20}$O as a function of the distance of closest approach \cite{Cor11} in Fig. \ref{fig:P2_fct_D_O20_O20}.
We see that the pair transfer probability varies almost exponentially with the distance of closest approach.
A remarkable point is the absence of individual transfer.
The probability to transfer one neutron is smaller than $10^{-4}$ for all  the calculations of the reaction $^{20}$O + $^{20}$O which were carried out in this paper. 
This reaction is an extreme case where the transfer is only due to the Josephson effect. In Fig. \ref{fig:distr}, we compare the results obtained by using TDHFB with that using  TDHF. In the case of TDHF, the dominant transfer probability is the one-neutron, and the pair transfer probability is close to $P_1^2/4$ which is the expected value assuming uncorrelated particles \cite{Hag15}.

In the case with superfluidity, the pair transfer is purely simultaneous. 
The absence of individual neutron transfer may be due to the large negative Q-value for one-neutron transfer $Q_{1n}=-3.8$ MeV while the pair transfer Q-value is $Q_{2n}=-0.9$ MeV. 
Here, the Q-values are computed from the experimental masses because the determination of the theoretical Q-value is complicated by the need to compute nuclei with odd-particle number.

\begin{figure}[htb]
\begin{center}
\includegraphics[width= 0.9 \linewidth]{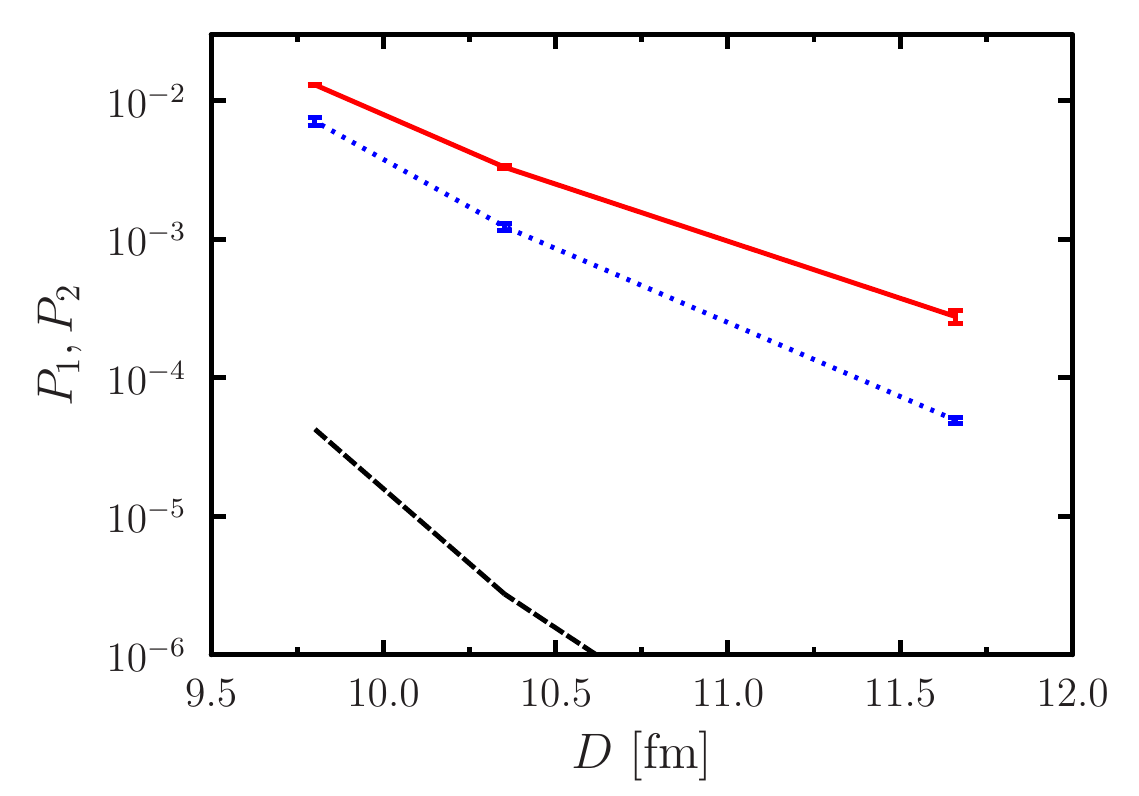}
\end{center}
\caption{ Transfer probability of one- (red solid line) and two-neutrons (blue dotted line)  as a function of the distance of closest approach for the reaction $^{14}$O+$^{20}$O. 
The dashed black line represents $P_1^2/4$.
}
\label{fig:P_fct_D_O14_O20}
\end{figure}

The second case studied is the reaction  $^{14}$O+$^{20}$O in which  the asymmetry of the reaction makes the transfer more favorable because of the natural tendency to equilibrate the masses of the fragments. 

We can see in Fig. \ref{fig:P_fct_D_O14_O20} the neutron transfer probabilities from the $^{20}$O fragments to the $^{14}$O. 
A large enhancement factor is found between the pair transfer probability and $P_1^2/4$. 
Note that it is also the expected value in the TDHF calculation without pairing \cite{Sca13}.
This enhancement of the pair transfer of about two orders of magnitude implies that in the present calculation the pairing correlations take a large role in the transfer process.

%\section{Summary and concluding remarks} 
%\label{Summary}

\textit{Summary.} The evolution of a set of HFB states is calculated assuming independent TDHFB evolution. 
The pair transfer probability is then computed from the mixed states obtained with the particle number projector. 
This method can be used to restore other symmetries like the initial spherical symmetry for reactions between deformed nuclei.
The method is successfully tested on a toy model where it can fairly reproduce the exact solution. 
The method is then applied to the realistic 3D TDHFB calculations with the Gogny force. 
Several prescriptions are proposed to obtain a correct behavior of the probability of particle transfer 
as a function of time after the reaction.
The calculation shows a large enhancement of the pair transfer probability. In the case of the $^{20}$O+$^{20}$O the single neutron transfer probability is completely suppressed and  in the $^{20}$O+$^{14}$O reaction a large enhancement factor arises.
With this result, we expect in future calculation to closely reproduce the experimental data for the sub-barrier transfer probabilities in reactions where there is large enhancement factor  \cite{Cor11,Mon14}. \\

\begin{acknowledgments}
 The authors thank the members of the DFT meeting for the exciting discussions.
The authors thank Professors T.~Nakatsukasa
for discussions and comments. 
The authors thank   D. Lacroix and D. Regnier for discussion and for pointing out a mistake in the calculation of the solution of the toy model.
This research work is partly supported by results 
of HPCI Systems Research Projects (Project ID hp170068). 
Using COMA at the CCS in University of Tsukuba, the numerical calculations 
were partly performed.  
\end{acknowledgments}


\begin{thebibliography}{99}



\bibitem{Ave08} B. Avez, C. Simenel, and Ph. Chomaz, Phys. Rev. C {\bf 78}, 044318 (2008).

\bibitem{Eba10} {S. Ebata, T. Nakatsukasa, T. Inakura, K. Yoshida, Y. Hashimoto, and K. Yabana, Phys. Rev. C {\bf 82}, 034306  (2010).}

\bibitem{Eba14} S. Ebata, T. Nakatsukasa, and T. Inakura, Phys. Rev. C {\bf 90}, 024303 (2014).

\bibitem{Sca13} G. Scamps, and D. Lacroix, Phys. Rev. C {\bf 87}, 014605 (2013).

\bibitem{Ste11} I. Stetcu, A. Bulgac, P. Magierski, and K. J. Roche, Phys. Rev. C {\bf 84}, 051309(R) (2011).


%5


\bibitem{Bul16} A. Bulgac, P. Magierski, K.J. Roche, and I. Stetcu, Phys. Rev. Lett. {\bf 116}, 122504 (2016).

\bibitem{Has12} Y. Hashimoto, Eur. Phys. J. A {\bf 48}, 1 (2012).

\bibitem{Has16} Y. Hashimoto, and G. Scamps, Phys. Rev. C {\bf 94}, 014610 (2016).

\bibitem{Toh02} M. Tohyama, and A.S. Umar, Phys. Lett. B {\bf 549}, 72-78 (2002).

\bibitem{Ass09} M. Assi\'e, and D. Lacroix, Phys. Rev. Lett. {\bf 102}, 202501 (2009).


%10


\bibitem{Toh16} M. Tohyama, and A.S. Umar, , Phys. Rev. C {\bf 93}, 034607 (2016).

\bibitem{Gou05} H. Goutte, J. F. Berger, P. Casoli, and D. Gogny, Phys. Rev. C {\bf 71}, 024316 (2005).

\bibitem{Reg16}  D. Regnier, N. Dubray, N. Schunck, and M. Verriere,  Phys. Rev. C  {\bf 93}, 054611 (2016).

\bibitem{Zde17} A. Zdeb, A. Dobrowolski, and M. Warda, Phys. Rev. C  {\bf 95}, 054608 (2017).

\bibitem{Uma06} A.S.Umar, and V.E.Oberacker, Phys.Rev.C {\bf 74}, 061601 (2006).

%15

\bibitem{Uma15}  A.S. Umar, V.E. Oberacker, and C. Simenel, Phys. Rev. C {\bf 92}, 024621 (2015).

\bibitem{bal84}  R. Balian, and M. Vénéroni, Phys. Lett. B {\bf 136}, 301 (1984).

\bibitem{bro08}  J. M. A. Broomfield, and P. D. Stevenson, J. Phys. G {\bf 35}, 095102 (2008).

\bibitem{Sim11BV}  C. Simenel, Phys. Rev. Lett. {\bf 106}, 112502 (2011).

\bibitem{sim12b}  C. Simenel, Eur. Phys. J. A {\bf 48}, 152 (2012).

%20

\bibitem{Ayi08} S. Ayik,  Phys. Lett. B {\bf 658}, 174 (2008).

\bibitem{Ayi09} S. Ayik, K. Washiyama, and D. Lacroix, Phys. Rev. C {\bf 79}, 054606 (2009).

\bibitem{Was09} K. Washiyama, S. Ayik, and D. Lacroix, Phys. Rev. C {\bf 80}, 031602(R) (2009).

\bibitem{Yil11} B. Yilmaz, S. Ayik, D. Lacroix, and K. Washiyama, Phys. Rev. C {\bf 83}, 064615 (2011).

\bibitem{Tan17} Y. Tanimura, D. Lacroix, and S. Ayik, Phys. Rev. Lett. {\bf 118}, 152501 (2017).

%25


\bibitem{Mag16} P. Magierski, K. Sekizawa, and G. Wlazlowski, Phys. Rev. Lett. {\bf 119}, 042501 (2017).

\bibitem{Sek17}  K. Sekizawa, G. Wlazłowski, and P. Magierski, arXiv:1705.04902 (2017).

\bibitem{Bul17} A. Bulgac and S. Jin,  Phys. Rev. Lett.  {\bf 119}, 052501 (2017).

\bibitem{Sim10} C. Simenel, Phys. Rev. Lett. {\bf 105}, 192701 (2010).

\bibitem{Sek13} K. Sekizawa, and K. Yabana, Phys. Rev. C {\bf 88}, 014614 (2013); Phys. Rev. C {\bf 90}, 064614 (2014). 

%30

\bibitem{Son15} Sonika, et al., Phys. Rev. C {\bf 92}, 024603 (2015).

\bibitem{Sca17} G. Scamps, C. Rodriguez-Tajes, D. Lacroix, and F. Farget, Phys. Rev. C  {\bf 95}, 024613 (2017).

\bibitem{Fom70} V. N. Fomenko, J. Phys. G {\bf 3}, 8 (1970).

\bibitem{Sim14}  C. Simenel, J. Phys. G: Nucl. Part. Phys. {\bf 41}, 094007 (2014).

\bibitem{Sca17_proc} G. Scamps, Y. Hashimoto,  arXiv:1704.00930 (2017).

%35

\bibitem{Ber12}  G.F. Bertsch, L.M. Robledo, Phys. Rev. Lett. {\bf 108}, 042505 (2012).

\bibitem{RS} P. Ring and P. Schuck, {\it The Nuclear Many-Body Problem}.

\bibitem{Cor11} L. Corradi, et al., Phys. Rev. C {\bf 84}, 034603, (2011).

\bibitem{Hag15} K. Hagino, and G. Scamps, Phys. Rev. C {\bf 92} 064602  (2015).

\bibitem{Mon14} D. Montanari, et al., Phys. Rev. Lett. {\bf 113}, 052501 (2014).

%40
 



\end{thebibliography}
\end{document}